\documentclass[11pt]{article}
\usepackage{amsmath}
\usepackage{amssymb}
\usepackage{epsfig}
\usepackage[latin5]{inputenc}

\numberwithin{equation}{section}

\newcommand{\dr}{\partial_{\rho}}

\newcommand{\ii}{i}

\DeclareMathOperator{\Sl}{sl} 
 
\DeclareMathOperator{\SL}{SL}

\newtheorem{prop}{Proposition}
\newtheorem{thm}{Theorem}

\begin{document}
\title{\bf
\Large On Some Canonical Classes of Cubic-Quintic Nonlinear Schr\"{o}dinger Equations }

\author{C. \"{O}zemir\footnotemark[1]\thanks{Department of Mathematics, Faculty of Science and Letters,
Istanbul Technical University, 34469 Istanbul,
Turkey, e-mail: ozemir@itu.edu.tr}
}

\date{\today}

\maketitle

\begin{abstract}
In this paper we bring into attention   variable coefficient  cubic-quintic nonlinear Schr\"{o}dinger   equations which admit Lie symmetry algebras of dimension four. Within this family, we obtain the reductions of canonical equations of  nonequivalent classes to ordinary differential equations using tools of Lie theory. Painlevé integrability of these reduced equations is investigated. Exact solutions through truncated Painlevé expansions are achieved in some cases. One of these solutions, a conformal-group invariant one, exhibits blow-up behaviour in finite time in $L_p$, $L_\infty$ norm and in distributional sense.
\end{abstract}

\section{Introduction}
\normalsize In this paper we aim at studying  a class of
cubic-quintic nonlinear Schr\"{o}dinger (CQNLS) equations, given in the form

\begin{equation}\label{canonic}
iu_t+u_{xx}+g(x,t)|u|^2u+q(x,t)|u|^4u+h(x,t)u=0
\end{equation}
in which  the complex coefficients $g,q$ and $h$ will be assumed to have some specific forms so that the equation under consideration admits a four-dimensional Lie symmetry algebra.

The motivation for this study comes from the recent work \cite{OzemirGungor2013}   on  a general class of  cubic-quintic nonlinear Schr\"{o}dinger equations given as
\begin{equation}\label{cqsch}
iu_t+f(x,t)u_{xx}+k(x,t)\,u_x+g(x,t)|u|^2u+q(x,t)|u|^4u+h(x,t)u=0.
\end{equation}
We performed classification of this family of equations according to Lie symmetry algebras they can admit. There $u$ is a complex-valued function, $f$ is real-valued, $k, g, q, h$ are complex-valued functions of the form $k=k_1(x,t)+i\,k_2(x,t)$, $g(x,t)=g_1(x,t)+ig_2(x,t)$, $q(x,t)=q_1(x,t)+iq_2(x,t)$ and
$h(x,t)=h_1(x,t)+ih_2(x,t)$ with the assumption that   $g\not\equiv 0$ or $q\not\equiv 0$, that is,
at least one of $g_1,g_2,q_1,q_2$ is different from zero. Eq. \eqref{cqsch} contains two physically important equations: cubic Schr\"{o}dinger equation for $k=q=0$ and quintic Schr\"{o}dinger equation for $k=g=0$ in one space dimension.

In \cite{OzemirGungor2013} we transformed \eqref{cqsch} to \eqref{canonic} by point transformations. Therefore \eqref{canonic} appears as a canonical equation
when classifying the family \eqref{cqsch} with respect to Lie symmetries they can admit. We would like to mention main results of this paper. We showed that the symmetry algebra $L$ of Eq. \eqref{canonic} (equivalently, of Eq. \eqref{cqsch}) is at most six-dimensional, that is, $1\leq \dim L \leq6$.  The following results concern the canonical equation   \eqref{canonic}, therefore they actually stand for the (in look) more general family \eqref{cqsch}.

\begin{itemize}
\item[R1.]  Any CQNLS equation within class \eqref{canonic}  having a 6-dimensional symmetry algebra is equivalent to the quintic constant-coefficient equation with $q=q_1+iq_2$, $g=h=0$.
\item[R2.] Any CQNLS equation within class \eqref{canonic}  having a 5-dimensional symmetry algebra is equivalent to the cubic constant-coefficient equation with $g=g_1+ig_2$, $q=h=0$.
\item[R3.] The symmetry algebra of the genuine (g and q not both zero) variable
coefficient CQNLS equation can be at most 4-dimensional. There are
precisely four inequivalent classes of equations in this case.

\item[R4.] None of these classes can be transformed to the standard constant-coefficient cubic-quintic equation with $g=g_1+ig_2$, $q=q_1+iq_2$, $h=0$.
\end{itemize}
According to these results, when a variable coefficient CQNLS equation has a  5- or 6-dimensional Lie symmetry algebra, it can be transformed to well-known constant coefficient cubic or quintic NLS equations, respectively. In the case when a variable coefficient CQNLS equation has a 4-dimensional symmetry algebra, we encountered four different classes of equations summarized in the following table.
\begin{table}[ht]
\caption{Four dimensional symmetry algebras and the coefficients in \eqref{canonic}} 
\centering 
\begin{tabular}{c c c c c } 
\hline\hline 
No & Algebra  &  $g$  & $q$  &  $h$ \\ [0.5ex] 
\hline 
$L_1$ & $\{T,D_1,C_1,W\}$                 & $(g_1+ig_2)\frac{1}{x}$    & $(q_1+iq_2)$         & $(h_1+ih_2)\frac{1}{x^2}$   \\
$L_2$ & $\{T,P,B,W\}$                     & $(g_1+ig_2)$               & $(q_1+iq_2)$         & $ih_2$                      \\
$L_3$ & $\{P,B,D_2,W\}$                   & $(g_1+ig_2)$               & $(q_1+iq_2)t$        & $i\frac{h_2}{t}$                  \\
$L_4$ & $\{P, B,C_2,W\}$                  & $(g_1+ig_2)$               & $(q_1+iq_2)(1+t^2)$  & $i\frac{t+h_2}{2(1+t^2)}$   \\ [1ex] 
\hline 
\end{tabular}
\label{algebras} 
\end{table}

The basis elements of subalgebras are given as follows:
\begin{align}
&T=\partial_{t},  \quad P=\partial_{x}, \quad
W=\partial_{\omega}, \quad
B=t\partial_{x}+\frac{1}{2}x\partial_{\omega}, \label{T}\\
&D_1=x\partial_x+2t\partial_t-\frac{1}{2}\rho\partial_{\rho}, \quad D_2=\frac{1}{2}x\partial_x+t\partial_t-\frac{1}{2}\rho\partial_{\rho},\\
&C_1=xt\partial_x+t^2\partial_t-\frac{1}{2}t\rho\partial_{\rho}+\frac{1}{4}x^2\partial_{\omega},\quad C_2=xt\partial_x+(1+t^2)\partial_t-t\rho\dr+\frac{1}{4}x^2\partial_{\omega}. \label{C1}
\end{align}
Here $u:\mathbb{R}^2\rightarrow \mathbb{C}$ is expressed in terms
of the modulus and the phase
\begin{equation}\label{psi}
u(x,t)=\rho(x,t)e^{i \omega(x,t)}.
\end{equation}
Equations \eqref{canonic} with coefficients from Table 1 are representatives of class of equations which admit non-isomorphic four-dimensional symmetry algebras. If we are to say that the symmetry algebras do not extend to five- or six-dimensional ones at all, we have to put the conditions $(g,h)\neq (0,0)$ for $L_1$, $(g,h_2)\neq (0,0)$ or $(q,h_2)\neq (0,0)$  for $L_2$, $(g,h_2)\neq (0,\frac{1}{4})$ or $(q,h_2)\neq (0,\{0,\frac{1}{2}\})$ for $L_3$ and $(g,h_2)\neq (0,0)$ for $L_4$ \cite{OzemirGungor2013}. Let us note that $L_1$ is non-solvable, $L_2$ is nilpotent, $L_3$ and $L_4$ are  solvable algebras.  Besides, $L_1$ is  decomposable whereas the others are not. Eq. \eqref{canonic} with coefficients from Table 1 cannot be transformed to a constant-coefficient equation, therefore they will be the main subject of this article.

Before moving to obtaining reduced equations for each representative equation for algebras $L_1,L_2,L_3,L_4$, let us mention another noteworthy point for which the results of this paper may be of use. Suppose one is interested in finding radial solutions to
\begin{equation}\label{neqn}
iu_t+\Delta u +g(x,t)|u|^2u+q(x,t)|u|^4u+h(x,t)u=0, \quad (x,t)\in \mathbb{R}^{N+1}\times \mathbb{R}.
\end{equation}
Radial solutions of \eqref{neqn} satisfy
\begin{equation}\label{radial}
iu_t+u_{xx}+\frac{N}{x}u_x+g(x,t)|u|^2u+q(x,t)|u|^4u+h(x,t)u=0, \quad x>0
\end{equation}
where $x$ plays the role of the radial coordinate. $u=x^{N/2}\tilde u$ transforms \eqref{canonic} to \eqref{radial} (up to a change in the coefficients). Thus using the results of this paper one can easily produce the corresponding results for the family \eqref{radial} and hence for the multidimensional one \eqref{neqn}.
\section{Reductions Through One-Dimensional Subalgebras}
Now we proceed to find the reduced ODEs for any representative equation \eqref{canonic} for $L_1,L_2,L_3,L_4$ with coefficients from Table 1.  Complex equation \eqref{canonic} has a four-dimensional solution space with two dependent and two independent variables. In order to reduce the solution space to a three-dimensional one with two dependent and one independent variable (i.e., a system of ODEs), we only need one-dimensional subalgebras of $L_1,L_2,L_3,L_4$. Thus we  restrict ourselves to subalgebras of each symmetry algebra having generic orbits of codimension 3 in the solution space.

The classification of one-dimensional subalgebras  under the
action of the group of inner automorphisms of the four-dimensional
symmetry groups is a standard one. We do not provide the
calculations leading to the conjugacy inequivalent list of
subalgebras. The classification method   can be found  for example
in \cite{Olver1991,Ovsiannikov1982,Winternitz1992}. We can refer to \cite{PateraWinternitz1977} for the classification of subalgebras of real three- and four-dimensional Lie algebras, from which the content of the following Table 2 is a direct consequence.
\begin{table}[ht]\label{t1}
\caption{One-dim. subalgebras of 4-dim. algebras under  the adjoint action of the full symmetry group} 
\centering 
\begin{tabular}{l l l l  } 
\hline\hline 
Algebra  &   &  Subalgebra  & $a,b,c\in\mathbb{R}$, $\epsilon=\mp1$   \\ [0.5ex] 
\hline 
$L_1$ & $L_{1.1}=\{T+C_1+aW\}$         & $L_{1.2}=\{D_1+bW\}$ & $L_{1.3}=\{T+cW\}$  \\ 
\hline
$L_2$ & $L_{2.1}=\{P\}$              & $L_{2.2}=\{T+aW\}$ & $L_{2.3}=\{B+bT\}$  \\
      & $L_{2.4}=\{W\}$              &                    &                      \\
\hline
$L_3$ & $L_{3.1}=\{P\}$              & $L_{3.2}=\{B\}$    & $L_{3.3}=\{P+\epsilon B\}$    \\
      &$L_{3.4}=\{D_2+aW\}$          &  $L_{3.5}=\{W\}$   &                                \\
\hline
$L_4$       & $L_{4.1}=\{B\}$    &$L_{4.2}=\{C_2+aW\}$      & $L_{4.3}=\{W\}$   \\           \hline 
\end{tabular}
\label{table:nonlin} 
\end{table}

In view of the formulation \eqref{psi} we express \eqref{canonic}
as a system of two real equations for the modulus $\rho$ and the
phase $\omega$ as follows
\begin{subequations}\label{reduce}
\begin{eqnarray}
   \label{reduce1}  -\rho \, \omega_t+\rho_{xx}-\rho \omega_x^2+g_1\rho^3+q_1\rho^5+h_1\rho&=&0, \\
   \label{reduce2}  \rho_t+2\rho_x \omega_x+\rho\omega_{xx}+g_2\rho^3+q_2\rho^5+h_2\rho&=&0.
\end{eqnarray}
\end{subequations}
\subsection{Non-solvable algebra ${L_1=\{T,D_1,C_1,W\}}$}
Let us note that the algebra has the direct sum structure
$L_1=\Sl(2,\mathbb{R})\oplus \{W\}$ and the representative
equation of the algebra is
\begin{equation}\label{eqL1}
iu_t+u_{xx}+(g_1+ig_2 )\frac{1}{x} |u|^2 u+(q_1+iq_2)|u|^4
u+(h_1+i h_2)\frac{1}{x^2} u=0
\end{equation}
with the real constants $g_1,g_2,q_1,q_2,h_1, h_2$.

\vspace{\baselineskip} \noindent\emph{\textbf{Subalgebra
$\mathbf{{L_{1.1}}}$}} Invariance under the
subalgebra $L_{1.1}$ implies that the solution has the form
\begin{equation}\label{sol11}
    u(x,t)=\frac{M(\xi)}{\sqrt{x}} \exp\left[\ii\Big(a \arctan t+\frac{x^2
    t}{4(1+t^2)}+P(\xi)\Big)\right],\quad \xi=\frac{x^2}{1+t^2}
\end{equation}
 and the reduced system of equations satisfied by $M(\xi)$ and $P(\xi)$ are
\begin{subequations}\label{11ab}
\begin{eqnarray}
   \label{11a}  &&4 \xi^2 M''-4\xi^2M P'^2+\Big(\frac{3-\xi^2}{4}-a \xi+h_1\Big)M+g_1 M^3+q_1M^5=0, \\
   \label{11b}  &&4 \xi^2 (M^2 P')'+h_2M^2+g_2M^4+q_2M^6=0.
\end{eqnarray}
\end{subequations}
We first need to decouple these equations to solve for the
functions $M$ and $P$. It is seen that an integral of \eqref{11b}
can be obtained for two different cases of the constants.

\noindent (i) $g_2=q_2=0$, $h_2\neq0$

It can be shown that the system \eqref{11ab} amounts to
integrating a third order nonlinear ordinary differential equation
from \eqref{11a}:
\begin{equation}\label{113}
Y'Y'''-\frac{1}{2}Y''^2+\frac{2}{\xi}Y'Y''+\frac{1}{2\xi^2}\Big(\frac{3-\xi^2}{4}-a
\xi+h_1\Big)Y'^2-\frac{2g_1}{h_2}Y'^3+\frac{8q_1\xi^2}{h_2^2}Y'^4-\frac{h_2^2}{8\xi^4}(Y+C)^2=0,
\end{equation}
where the functions $M,P$ of \eqref{sol11} are related to $Y(\xi)$
by the relations
\begin{equation}
M(\xi)=2\xi\sqrt{-\frac{1}{h_2} Y'}\,, \qquad
P(\xi)=-\frac{h_2}{4}\int \frac{Y+C}{\xi^2 Y'} \, d\xi.
\end{equation}
\noindent (ii) $g_2=q_2=h_2=0$

In this case we can easily decouple the reduced system of
equations. Integration of  \eqref{11b} gives
\begin{equation}
M^2P'=C,  \qquad P(\xi)=\int \frac{C}{M^2}\,d\xi, \quad
C=\rm{const.}
\end{equation}
and from  \eqref{11a} we obtain the equation for $M$
\begin{equation}\label{112}
M''=-\frac{q_1}{4\xi^2}M^5-\frac{g_1}{4\xi^2}M^3+\Big(\frac{1}{16}+\frac{a}{4\xi}-\frac{3+4h_1}{16\xi^2}\Big)M+C^2M^{-3}.
\end{equation}
For the remaining subalgebras, we only state the results obtained
through similar operations.

\vspace{\baselineskip} \noindent\emph{\textbf{Subalgebra
$\mathbf{{L_{1.2}}}$}}
\begin{equation}\label{sol12}
   u(x,t)=\frac{M(\xi)}{\sqrt{x}} \exp\Big[\ii\big(b\ln x+P(\xi)\big)\Big],\quad
    \xi=\frac{x^2}{t}
\end{equation}
results in the system
\begin{subequations}\label{12ab}
\begin{eqnarray}
   \label{12a}  &4 \xi^2 M''-4\xi^2MP'^2+\xi(\xi-4b)M P'+(\frac{3}{4}+h_1-b^2)M+g_1 M^3+q_1M^5=0,\quad  \\
   \label{12b}  &\xi^2\left[M^2\Big(\frac{2b}{\xi}-\frac{1}{2}+4P'\Big)\right]'+h_2M^2+g_2 M^4+q_2M^6=0.
\end{eqnarray}
\end{subequations}
\noindent (i)  $g_2=q_2=0$, $h_2\neq0$
\begin{subequations}
\begin{eqnarray}
   \label{123}  && Y'Y'''-\frac{1}{2}Y''^2+\frac{2}{\xi}Y'Y''+\Big(\frac{3+4h_1}{8\xi^2}-\frac{b}{4\xi}+\frac{1}{32}\Big)Y'^2-\frac{g_1}{2h_2}Y'^3+\frac{q_1\xi^2}{2h_2^2}Y'^4 \nonumber \\
   \label{123}  && -\frac{h_2^2}{8\xi^4}(Y+C)^2=0, \\
                &&M^2=-\frac{1}{h_2}\,\xi^2\,Y', \qquad P=\frac{\xi}{8}-\frac{b}{2}\ln \xi-\frac{h_2}{4}\int
   \frac{Y+C}{\xi^2Y'}\,d\xi,
\end{eqnarray}
\end{subequations}
\noindent (ii)  $g_2=q_2=h_2=0$
\begin{subequations}
\begin{eqnarray}
   \label{122}  && M''=C^2 M^{-3}+(\frac{b}{8\xi}-\frac{3+4h_1}{16\xi^2}-\frac{1}{64})M-\frac{g_1}{4\xi^2}M^3-\frac{q_1}{4\xi^2}M^5,\\
                &&P=\frac{\xi}{8}-\frac{b}{2}\ln \xi-C\int
   \frac{d\xi}{M^2}.
\end{eqnarray}
\end{subequations}

\vspace{\baselineskip} \noindent\emph{\textbf{Subalgebra
$\mathbf{{L_{1.3}}}$}}
\begin{equation}\label{sol13}
  u(x,t)=M(x) \exp\Big[\ii\big(ct+P(x)\big)\Big]
\end{equation}
results in the system
\begin{subequations}\label{13ab}
\begin{eqnarray}
   \label{13a}  &&M''-MP'^2+(\frac{h_1}{x^2}-c)M+\frac{g_1}{x}M^3+q_1M^5=0, \\
   \label{13b}  &&x^2(M^2P')'+h_2M^2+g_2xM^4+q_2x^2M^6=0.
\end{eqnarray}
\end{subequations}
\noindent (i)  $g_2=q_2=0$, $h_2\neq0$
\begin{subequations}
\begin{eqnarray}
     &&  Y' Y'''-\frac{1}{2}Y''^2+\frac{2}{x} Y' Y''+2\Big(\frac{h_1}{x^2}-c\Big) Y'^2-\frac{2 g_1 x}{h_2}Y'^3+\frac{2 q_1 x^4}{h_2^2}Y'^4\nonumber \\
 \label{133}  &&-\frac{2h_2^2}{x^4}(Y+C)^2=0,\\
                &&M(x)=\left(-\frac{1}{h_2} x^2Y'\right)^{1/2}, \qquad P(x)=-h_2
    \int \frac{Y+C}{x^2 Y'}\,dx.
\end{eqnarray}
\end{subequations}
\noindent (ii)  $g_2=q_2=h_2=0$
\begin{subequations}
\begin{eqnarray}
   \label{132}  && M''=C^2M^{-3}+\Big(c-\frac{h_1}{x^2}\Big)M-\frac{g_1}{x}M^3-q_1M^5,\\
                &&P(x)=\int \frac{C}{M^2}\, dx.
\end{eqnarray}
\end{subequations}
\subsection{Nilpotent algebra ${L_2=\{T,P,B,W\}}$}
The canonical equation has the form
\begin{equation}\label{eqL2}
iu_t+u_{xx}+(g_1+i g_2 )|u|^2 u+(q_1+iq_2)|u|^4 u+i h_2 u=0.
\end{equation}

\vspace{\baselineskip} \noindent\emph{\textbf{Subalgebra
$\mathbf{{L_{2.1}}}$}} The invariant solution $u(x,t)=M(t)
\exp\big(\ii P(t)\big)$ leads to the first-order system
\begin{equation}\label{211}
P'=g_1M^2+q_1M^4, \qquad  M'+h_2M+g_2M^3+q_2M^5=0
\end{equation}
which is straightforward to integrate upon given values of the
constants.

\vspace{\baselineskip} \noindent\emph{\textbf{Subalgebra
$\mathbf{{L_{2.2}}}$}} $u(x,t)=M(x) \exp\Big[\ii
\big(at+P(x)\big)\Big]$ leads to the system
\begin{subequations}\label{22ab}
\begin{eqnarray}
   \label{22a}  &&M''-MP'^2-aM+g_1M^3+q_1M^5=0,\\
   \label{22b}  &&(M^2P')'+h_2M^2+g_2M^4+q_2M^6=0.
\end{eqnarray}
\end{subequations}
\noindent (i)  $g_2=q_2=0$, $h_2\neq0$
\begin{subequations}
\begin{eqnarray}
   \label{223}  &&  Y'Y'''-\frac{1}{2}Y''^2-2 a Y'^2-\frac{2g_1}{h_2} Y'^3+\frac{2q_1}{h_2^2}Y'^4-2
     h_2^2(Y+C)^2=0,\\
                &&M(x)=\left(-\frac{1}{h_2}Y'\right)^{1/2}, \qquad P(x)=-h_2
    \int \frac{Y+C}{ Y'}\,dx.
\end{eqnarray}
\end{subequations}
\noindent (ii)  $g_2=q_2=h_2=0$
\begin{subequations}
\begin{eqnarray}
   \label{222}  && M''=C^2M^{-3}+a M-g_1M^3-q_1M^5,\\
                &&P(x)=\int \frac{C}{M^2}\, dx.
\end{eqnarray}
\end{subequations}

\vspace{\baselineskip} \noindent\emph{\textbf{Subalgebra
$\mathbf{{L_{2.3}}}$}}

\noindent (A) The case $b\neq0$.  The invariant solution
\begin{equation}\label{sol23}
    u(x,t)=M(\xi) \exp\Big[\ii \Big(\frac{1}{2 b}x t-\frac{1}{6
    b^2}t^3+P(\xi)\Big)\Big], \qquad \xi=b x-\frac{t^2}{2}
\end{equation}
gives the system
\begin{subequations} \label{23ab}
\begin{eqnarray}
   \label{23a}  &&M''-MP'^2-\frac{\xi}{2b^4}M+\frac{g_1}{b^2}M^3+\frac{q_1}{b^2}M^5=0,\\
   \label{23b}  &&b^2(M^2P')'+h_2M^2+g_2M^4+q_2M^6=0.
\end{eqnarray}
\end{subequations}
\noindent (i)  $g_2=q_2=0$, $h_2\neq0$
\begin{subequations}
\begin{eqnarray}
   \label{233}  &&  Y'Y'''-\frac{1}{2}Y''^2-\frac{\xi}{b^4} Y'^2-\frac{2g_1}{h_2} Y'^3+\frac{2q_1b^2}{h_2^2} Y'^4-
     \frac{2 h_2^2}{b^4}(Y+C)^2=0,\\
                && M(\xi)=(-\frac{b^2}{h_2}Y')^{1/2}, \qquad P(\xi)=-\frac{h_2}{b^2} \int \frac{Y+C}{Y'} \, d\xi.
\end{eqnarray}
\end{subequations}
\noindent (ii)  $g_2=q_2=h_2=0$
\begin{subequations}
\begin{eqnarray}
   \label{232}  && M''=C^2M^{-3}+\frac{\xi}{2b^4}M-\frac{g_1}{b^2}M^3-\frac{q_1}{b^2}M^5,\\
                &&P(x)=\int \frac{C}{M^2}\, d\xi.
\end{eqnarray}
\end{subequations}
\noindent (B) The case $b=0$.
\begin{equation}\label{}
    u(x,t)=M(t) \exp\Big[\ii \Big(\frac{x^2}{4t}+P(t)\Big)\Big]
\end{equation}
is the form of the group-invariant solution and the reduced system
of equations is
\begin{equation}\label{231}
    P'=g_1M^2+q_1M^4,\qquad
    M'+(h_2+\frac{1}{2t}) M +g_2M^3+q_2M^5=0.
\end{equation}
This system is readily solved by standard methods.

\subsection{Solvable algebra ${L_3}=\{P,B,D_2,W\}$}
We note that the algebra is not decomposable and   the
representative equation  from Table 1 is
\begin{equation}\label{eqL3}
iu_t+u_{xx}+(g_1+ig_2 )|u|^2 u+(q_1+iq_2)t|u|^4 u+i \frac{h_2}{t}
u=0.
\end{equation}

\vspace{\baselineskip} \noindent\emph{\textbf{Subalgebra
$\mathbf{{L_{3.1}}}$}} $u(x,t)=M(t) \exp\big(\ii P(t)\big)$
leads to the  system
\begin{equation}\label{311}
P'=g_1M^2+q_1tM^4,  \quad M'+\frac{h_2}{t}M+g_2M^3+q_2tM^5=0.
\end{equation}
\vspace{\baselineskip} \noindent\emph{\textbf{Subalgebra
$\mathbf{{L_{3.2}}}$}} $u(x,t)=M(t)\exp\Big[\ii
\Big(\frac{x^2}{4t}+P(t)\Big)\Big]$ gives the reduction
\begin{equation}\label{321}
P'=g_1M^2+q_1tM^4, \quad M'+\frac{2h_2+1}{2t}M+g_2M^3+q_2tM^5=0.
\end{equation}
\vspace{\baselineskip} \noindent\emph{\textbf{Subalgebra
$\mathbf{{L_{3.3}}}$}}  $u(x,t)=M(t)\exp\Big[\ii
\Big(\frac{\epsilon x^2}{4(1+\epsilon t)}+P(t)\Big)\Big]$
gives
\begin{equation}\label{331}
P'=g_1M^2+q_1tM^4,\qquad
M'+\left(\frac{h_2}{t}+\frac{\epsilon}{2(1+\epsilon
t)}\right)M+g_2M^3+q_2tM^5=0.
\end{equation}
\vspace{\baselineskip} \noindent\emph{\textbf{Subalgebra
$\mathbf{{L_{3.4}}}$}} A group-invariant solution
invariant under  $L_{3.4}$ will be of the form
\begin{equation}\label{sol34}
u(x,t)=\frac{1}{x}M(\xi)\exp\Big[\ii \big(2a\ln
x+P(\xi)\big)\Big], \quad \xi=\frac{x^2}{t}
\end{equation}
where  $M,P$ will be solutions of the system
\begin{subequations}\label{34ab}
\begin{eqnarray}
   \label{34ai}       4\xi^2M''-4\xi^2MP'^2+\xi(\xi-8a)M P'-2\xi M'+2(1-2a^2)M+g_1M^3+\frac{q_1}{\xi}M^5=0,\quad \\
   \label{34bi}       4\xi^2(M^2P')'-2\xi M^2P'-\xi(\xi-8a)M M'+(h_2\xi-6a)M^2+g_2M^4+\frac{q_2}{\xi}M^6=0. \quad
\end{eqnarray}
\end{subequations}
\eqref{34bi} is satisfied if $g_2=q_2=0$, $h_2=1/4$,
$\displaystyle P(\xi)=\frac{\xi}{8}-a \ln \xi+P_0$ and \eqref{34ai} reduces to
\begin{equation}\label{342}
M''=\frac{1}{2\xi}M'+(\frac{a}{4\xi}-\frac{1}{2\xi^2}-\frac{1}{64})M-\frac{g_1}{4\xi^2}M^3-\frac{q_1}{4\xi^3}M^5.
\end{equation}
\subsection{Solvable algebra ${L_{4}=\{P,B,C_2,W\}}$}
The  last canonical equation under investigation will be
\begin{equation}\label{eqL4}
iu_t+u_{xx}+(g_1+ig_2)\, |u|^2 u+(q_1+iq_2)(1+t^2)|u|^4
u+i\frac{t+h_2}{2(1+t^2)} \,u=0.
\end{equation}
\vspace{\baselineskip} \noindent\emph{\textbf{Subalgebra
$\mathbf{{L_{4.1}}}$}}
$u(x,t)=M(t)\exp\Big[\ii\big(\frac{x^2}{4t}+P(t)\big)\Big]$ gives
 the  system
\begin{equation}\label{411}
P'=g_1M^2+q_1(1+t^2)M^4,  \quad
M'+\frac{1+h_2t+2t^2}{2t(1+t^2)}M+g_2M^3+q_2(1+t^2)M^5=0.
\end{equation}
\vspace{\baselineskip} \noindent\emph{\textbf{Subalgebra
$\mathbf{{L_{4.2}}}$}}
\begin{equation}\label{sol42}
u(x,t)=\frac{1}{\sqrt{1+t^2}}\,M(\xi) \exp\Big[\ii \Big(a \arctan
t+\frac{x^2t}{4(1+t^2)}+P(\xi)\Big)\Big], \qquad
\xi=\frac{x}{\sqrt{1+t^2}}
\end{equation}
results in the system
\begin{subequations}\label{42ab}
\begin{eqnarray}
   \label{42a}  &&M''-MP'^2-(a+\frac{\xi^2}{4})M+g_1M^3+q_1M^5=0, \\
   \label{42b}  &&(M^2P')'+\frac{h_2}{2}M^2+g_2M^4+q_2M^6=0.
\end{eqnarray}
\end{subequations}
\noindent (i)  $g_2=q_2=0$, $h_2\neq0$
\begin{subequations}
\begin{eqnarray}
   \label{423}  &&  Y'Y'''-\frac{1}{2}Y''^2-(\frac{\xi^2}{2}+2a) Y'^2-\frac{4g_1}{h_2} Y'^3+\frac{8q_1}{h_2^2}Y'^4-
     \frac{h_2^2}{2}(Y+C)^2=0,\\
                &&M(\xi)=\left(-\frac{2}{h_2}Y'\right)^{1/2}, \qquad P(\xi)=-\frac{h_2}{2}
    \int \frac{Y+C}{ Y'}\,d\xi.
\end{eqnarray}
\end{subequations}

\noindent (ii)  $g_2=q_2=h_2=0$
\begin{subequations}
\begin{eqnarray}
   \label{422}  && M''=C^2M^{-3}+\Big(a+\frac{\xi^2}{4}\Big)M-g_1M^3-q_1M^5,\\
                &&P(\xi)=\int \frac{C}{M^2}\, d\xi.
\end{eqnarray}
\end{subequations}

\section{Analysis of the Reduced Equations}
In this section we will analyze the reduced equations obtained in the previous section. We wonder whether any of those reduced equations are integrable, so as to obtain exact solutions for them, therefore to the PDEs they are originating from. The main tool for the search of integrability is going to be the Painlevé property of differential equations and related techniques based on this concept. We would like to mention a few lines about the methods we will utilize first.

 A differential equation is said to have the \emph{Painlevé property} if all its solutions do not include movable singularities other than poles. Painlevé himself classified all second order ordinary  differential equations in the complex plane which have this property \cite{PainleveI,PainleveII} and his work was completed by Gambier \cite{Gambier}. There are fifty canonical second order ODEs having this property. All but six of them are integrable in terms of known functions. The six canonical differential equations, named Painlevé DEs I-VI, define new transcendental functions. A complete list of these equations can be found in \cite{Ince}. This is the case for second-order ODEs, but not the end of the story. The classification was continued for higher order ODEs  afterwards (see for example \cite{CosgroveI,CosgroveII}). When we have a second- or higher-order nonlinear equation in hand integrate, or to look solutions for, checking any correspondence with these Painlevé-type equations is the very first station to stop by. This is going to be our first approach in analyzing the reduced equations.

In order to check whether a given partial differential equation has the Painlevé property,  a necessary condition was proposed in \cite{WTC}. According to the so-called WTC test, the  unknown function  $u(x,t)$ of a PDE (system) is expanded into a power series in the form of a Laurent series
\begin{equation}\label{}
u(x,t)=\sum_{j=0}^{\infty} \, u_j(x,t) \,\Phi(x,t)^{\alpha+j}.
\end{equation}
Here $\Phi(x,t)=0$ is an arbitrary singularity manifold (a curve in this case).  The power $\alpha$ is called the \emph{leading order}. If $\alpha$ assumes (negative) integer values, the Laurent series form excludes  singularities other than poles. In the case of a PDE system, upon plugging these form of expansions for dependent variables one obtains a system of the form $\displaystyle \mathbf{Q}(j)\mathbf{U}(j)=\mathbf{F}(j)$. It is not possible to determine coefficients $u_j$ for which $\det \mathbf{Q}(j)=0 $. They remain free, if this system is compatible for those indices $j_{res}$, called as \emph{resonances}. The partial differential equation is said to pass the Painlevé test if there arises the correct number of arbitrary functions as required by the Cauchy-Kovalewski theorem given by the expansion coefficients, and $\Phi(x,t)$ should be one of the arbitrary functions.

If a differential equation passes the P-test, it has a good chance of integrability and one can try transformation to the representative equations of the canonical classes mentioned above, besides making use of the results available in literature for them. If the consistency conditions at the resonance indices are not satisfied, the PDE does not pass the Painlevé test, therefore it does not have the Painlevé property. In this case, if one truncates the P-series at a finite term as
\begin{equation}\label{truncation}
u(x,t)=\sum_{j=0}^{N} \, u_j(x,t) \,\Phi(x,t)^{\alpha+j}
\end{equation}
and try to satisfy the consistency conditions for special cases of functions $u_j$ and $\Phi$ (which failed to remain free), \eqref{truncation} may turn out to be an exact analytical solution for the PDE. By this way one can also obtain B\"{a}cklund transformations relating solutions of the equation. We refer the interested reader to \cite{OG2006,Yomba1996,OG2011} and \cite{GHO2013}, among many other works on Schr\"{o}dinger and other type equations.

After this overview of the methods we shall make use of, we can turn back to have a look at the material we obtained in the previous section. Table \ref{numbers} is a guide to the equations obtained as
reductions for the canonical PDEs with symmetry algebras
$L_1,L_2,L_3,L_4$. In what follows, we list our findings about those reduced equations.
\begin{table}[h]
\caption{A summary of the reduced equations} 
\centering 
\begin{tabular}{c l} 
\hline\hline 
Order & Equation Number  \\ [0.5ex] 
\hline 1 & \eqref{211}, \eqref{231}, \eqref{311}, \eqref{321}, \eqref{331},  \eqref{411}   \\
\hline 2 &  \eqref{112}, \eqref{122}, \eqref{132}, \eqref{222}, \eqref{232}, \eqref{342}, \eqref{422} \\
\hline 3 & \eqref{113}, \eqref{123}, \eqref{133}, \eqref{223},
\eqref{233}, \eqref{423}  \\ \hline
\end{tabular}
\label{numbers} 
\end{table}
\begin{itemize}
\item The first order decoupled systems can be solved by elementary
methods, so we do not present any solution for them here.
\item Among the second-order equations, it was seen that
\eqref{222} passes the Painlevé test after the transformation $M=\sqrt{U}$ without any
restriction on the constants.  Let us note that, \eqref{222} is a
reduction for the PDE
\begin{equation}\label{g1q1}
iu_t+u_{xx}+g_1|u|^2u+q_1|u|^4u=0.
\end{equation}
Ref. \cite{Gagnon1989} is devoted to the search of exact solutions
of \eqref{g1q1}, and \eqref{222} is obtained as a direct
reduction, after all being completely analyzed to obtain elliptic
solutions. We refer the interested reader to this work. Note that
since \eqref{222} does not include the independent variable
explicitly, it can be integrated to a first-order equation by the
transformation $M'=W(M)$.
\item The second order equation \eqref{132} passes the Painlevé test after the transformation  $M=\sqrt{U}$ if
$g_1=h_1=0$ and the previous argument is also valid for this
equation.
\item Second-order reductions other than \eqref{132} and
\eqref{222} fail to pass the Painlevé test, therefore
they give pretty less expectation for integrability.
\item From the third-order equations, \eqref{223} does not include
the independent variable; this means it can be reduced to a
second-order equation by a transformation $Y'=W(Y)$.
\item All the third-order equations pass the Painlevé test if the variable $Y$ is replaced as $Y\rightarrow 1/F(\xi)$ or $Y\rightarrow 1/F'(\xi)$. After this change of variable, the third order equations convert to an equation in $F$ which has a leading behaviour with $\alpha=-1$ or $\alpha=-2$ with several branches successfully satisfying the conditions of the test.
\end{itemize}
As we have mentioned, among seven second order equations we obtained, only \eqref{132} and
\eqref{222} passed the Painlevé test under some restrictions on the parameters they include. However, under these conditions, the PDE \eqref{g1q1} they originated from turned out to be  one which was already analyzed in literature. Therefore, our first approach for analyzing the reduced equations,  which is to compare them with the representatives of P-integrable equations of second order did not give any new result to us.  At this moment we take another approach. We look for solutions through truncated P-expansions for our equations. Section 3.1 is an effort to do so for second-order reduced equations. Section 3.2 tries the same method not for the reduced equations but by directly attacking the canonical PDEs for the algebras given in Table 1.
\subsection{Exact solutions to second-order reduced equations through truncated expansions }
 All of the second-order equations listed in Table \ref{numbers} has a leading order $\alpha=-1/2$. This suggests that the transformation $M=\sqrt{U}$ will transform those equations to equations with dependent variable $U$, for which the P-series expansions will have a  leading-order  $\alpha=-1$.  For the moment let us denote the independent variables for second order equations in Table \ref{numbers} by $x$. The Painlevé series expansion  for solutions $U(x)$ will have the form
 \begin{equation}
 U(x)=\sum_{j=0}^{\infty}U_j(x)\Phi^{-1+j}(x).
 \end{equation}
We truncate this P-series expansion at the first term; that is, we propose a solution to our second-order ODEs in the form
\begin{equation}\label{subs}
 U(x)=\frac{U_0(x)}{\Phi(x)}.
\end{equation}
When we plug this  expression in the DE for $U$, terms which can be collected according to the powers $\Phi^{j}$, $j=-4,-3,-2,-1,0$ appear in the equation. By putting the condition that the coefficients of those $\Phi^j$s vanish, we try to determine the form of the functions $U_0(x)$ and $\Phi(x)$ and yet determine $U(x)$. For the seven second order ODEs in the Table \ref{numbers}, we have been successful in two cases.

\vspace{\baselineskip} \noindent\emph{\textbf{Subalgebra
$\mathbf{{L_{1.3}.}}$}} We start with Eq. \eqref{132}, including the constants $g_1,q_1,h_1$ of the original PDE, the constant $c$ arising from the group action and the integration constant $C$ obtained during the decoupling. Putting $M=\sqrt{U}$ in \eqref{132} gives
\begin{equation}\label{132u}
U''=\frac{1}{2U}\,U'^2+2(c-\frac{h_1}{x^2})U-\frac{2g_1}{x}U^2-2q_1U^3+\frac{2C^2}{U}.
\end{equation}
We impose \eqref{subs} in \eqref{132u} and require that coefficients of $\Phi^j$s arising vanish.  First of all, by the coefficient of $\Phi^{-4}$, it is necessary that  $q_1$ is a negative number, say $ q_1=-\beta^2$. Further calculations put several constraints on the constants included in the equation. We must have $c=C=0$. We see that it is possible to solve the overdetermined system of equations for $U_0$ and $\Phi$ in two different cases. If
$$g_1=\frac{2\gamma\beta}{\sqrt{3}}, \quad q_1=-\beta^2, \quad h_1=\frac{\gamma}{2}-\frac{\gamma^2}{4}, \quad \gamma \neq -1$$
we obtain
\begin{equation}\label{}
U_0(x)=\frac{\sqrt{3}k_1(1+\gamma)}{2\beta}\,x^\gamma, \qquad \Phi(x)=k_1x^{1+\gamma}+k_2
 \end{equation}
where $\beta$ and $\gamma$ are fixed by the given values of $g_1$, $q_1$ and $k_1,k_2$ are arbitrary constants. Similarly, if
$$g_1=\frac{-2\beta}{\sqrt{3}}, \quad q_1=-\beta^2, \quad h_1=-\frac{3}{4} $$
we find
\begin{equation}\label{}
U_0(x)=\frac{\sqrt{3}k_1}{2\beta}\,\frac{1}{x}, \qquad \Phi(x)=k_1\ln x+k_2.
\end{equation}
We state the final solution as follows.
\begin{prop} The canonical equation
\begin{equation}\label{eq}
iu_t+u_{xx}+\frac{g_1+ig_2}{x}|u|^2u+(q_1+iq_2)|u|^4u+\frac{h_1+ih_2}{x^2}u=0
\end{equation}
for the subalgebra $L_1=\mathrm{sl(2,\mathbb{R})}\oplus W$ is solved by the following functions.

\noindent(i). If $g_1=\frac{2\gamma\beta}{\sqrt{3}}$,  $q_1=-\beta^2$, $h_1=\frac{\gamma}{2}-\frac{\gamma^2}{4}$, $\gamma \neq -1$, $g_2=q_2=h_2=0:$
\begin{equation}\label{solprop1i}
u(x,t)=\zeta(x)=A\left[\frac{x^\gamma}{k_1x^{1+\gamma}+k_2}\right]^{1/2}e^{iP_0}, \qquad A^2=\frac{\sqrt{3}k_1(1+\gamma)}{2\beta}.
\end{equation}
(ii). If $g_1=\frac{-2\beta}{\sqrt{3}}$, $q_1=-\beta^2$, $h_1=-\frac{3}{4}$, $g_2=q_2=h_2=0:$
\begin{equation}
u(x,t)=\frac{Ae^{iP_0}}{[x(k_1\ln x+k_2)]^{1/2}}, \qquad A^2=\frac{\sqrt{3}k_1}{2\beta}.
\end{equation}
\end{prop}
\textbf{Remark 1.} In fact, \eqref{132} is of Painlevé type  if $g_1=h_1=0$. The above solution does not have such a requirement, therefore is a solution in non-Painlevé case.

\vspace{\baselineskip} \noindent\emph{\textbf{Subalgebra $\mathbf{{L_{2.2}.}}$}} Next we give the analysis for \eqref{222}. It includes the constants $g_1$ and $q_1$ originating from the reduced PDE, whereas $a$ is a free reduction parameter. Setting $M=\sqrt{U}$ results in
\begin{equation}\label{222u}
U''=\frac{1}{2U}\,U'^2+2aU-2g_1U^2-2q_1U^3+\frac{2C^2}{U}.
\end{equation}
We proceed by plugging \eqref{subs} in \eqref{222u}. Again it is necessary that $q_1=-\beta^2<0$.  We find that for
$$g_1=\frac{2\gamma\beta}{\sqrt{3}}, \quad q_1=-\beta^2, \quad a=\frac{\gamma^2}{4}$$
the functions $U_0$, $\Phi$ will be
\begin{equation}\label{}
U_0(x)=\frac{\sqrt{3}k_1\gamma}{2\beta}\,e^{\gamma x}, \qquad \Phi(x)=k_1e^{\gamma x}+k_2.
\end{equation}

\begin{prop} The canonical equation
\begin{equation}
iu_t+u_{xx}+(g_1+ig_2)|u|^2u+(q_1+iq_2)|u|^4u+ih_2u=0
\end{equation}
for the subalgebra $L_2$ is solved by
\begin{equation}
u(x,t)=A\left[\frac{e^{\gamma x}}{k_1 e^{\gamma x}+k_2}\right]^{1/2}e^{i \frac{\gamma^2 t}{4}}, \qquad A^2=\frac{\sqrt{3}k_1\gamma}{2\beta}
\end{equation}
where $g_1=\frac{2\gamma\beta}{\sqrt{3}}$, $q_1=-\beta^2$, $g_2=q_2=h_2=0$.
\end{prop}
\textbf{Remark 2.} One needs to keep in mind that \eqref{222} is a Painlevé-type equation without any restriction on the constants.

We tried the same procedure for all other second-order equations, arriving at failure.
\subsection{An exact solution by truncation in Painlevé series}
The method we are going to use is truncation in Painlevé series again. Unlike the previous subsection, this time we are focusing on a canonical PDE itself,
instead of its reductions.
We would like to note an exact solution of the variable
coefficient canonical equation
\begin{equation}\label{exact}
iu_t+u_{xx}+\frac{g_1+ig_2}{x}|u|^2u+(q_1+iq_2)|u|^4u+\frac{h_1+ih_2}{x^2}u=0
\end{equation}
of the symmetry algebra $L_1$.
For the case $q_1=q_2=0$, inspired by \cite{Yomba1996},  we obtained interesting exact solutions in
\cite{OG2011}, giving rise to blow-up analysis studied in \cite{GHO2013}.

In fact, when we make a leading order analysis to
\eqref{exact}, it is immediate to see that it cannot have the
Painlevé property as the balancing of the leading orders is not
possible because of the quintic term. Even the leading order $\alpha$ cannot be determined. However, we keep on tracking the route we followed
in the absence of the quintic term technically (see \cite{OG2011} for
details),  we write \eqref{exact} with its complex conjugate $v$ as
the system
\begin{subequations}\label{series}
\begin{eqnarray}
iu_t+u_{xx}+\frac{g_1+ig_2}{x}u^2v+(q_1+iq_2)u^3v^2+\frac{h_1+ih_2}{x^2}u=0,\\
-iv_t+v_{xx}+\frac{g_1-ig_2}{x}v^2u+(q_1-iq_2)v^3u^2+\frac{h_1-ih_2}{x^2}v=0
\end{eqnarray}
\end{subequations}
and propose a solution of the form
\begin{equation}\label{pro}
u(x,t)=u_0(x,t)\,\Phi^{-1-i\delta}(x,t), \quad
v(x,t)=v_0(x,t)\,\Phi^{-1+i\delta}(x,t).
\end{equation}
The system \eqref{series} has no leading order. Those powers $-1\mp i\delta$ are proposed in regard to cubic case studied in \cite{OG2011}. When we substitute \eqref{pro} in \eqref{series}, terms with
coefficients $\Phi^{-5\mp i\delta}$,$\Phi^{-3\mp
i\delta}$,$\Phi^{-2\mp i\delta}$,$\Phi^{-1\mp i\delta}$ appear in
the system. We shall note here first two of them:
\begin{subequations}\label{}
\begin{eqnarray}
(q_1+iq_2)u_0^3v_0^2\Phi^{-5- i\delta}+\Big[\frac{g_1+ig_2}{x}u_0^2 v_0+(2+3i\delta-\delta^2)u_0\Phi_x^2\Big]\Phi^{-3- i\delta},\\
(q_1-iq_2)u_0^2v_0^3\Phi^{-5+i\delta}+\Big[\frac{g_1-ig_2}{x}u_0
v_0^2+(2-3i\delta-\delta^2)v_0\Phi_x^2\Big]\Phi^{-3+ i\delta}.
\end{eqnarray}
\end{subequations}
In a usual truncation procedure one proceeds with requiring that
coefficients of $\Phi^{-5\mp i\delta}$ and $\Phi^{-3\mp i\delta}$
vanish separately. However, it is not possible to have that
coefficients of $\Phi^{-5\mp i\delta}$ vanish. We collected these
terms together in the form
\begin{subequations}\label{}
\begin{eqnarray}
\left[(q_1+iq_2)\Phi^{-2}\,u_0^2v_0^2+\frac{g_1+ig_2}{x}\,u_0 v_0+(2+3i\delta-\delta^2)\Phi_x^2\right]u_0\Phi^{-3- i\delta}=0,\\
\left[(q_1-iq_2)\Phi^{-2}\,u_0^2v_0^2+\frac{g_1-ig_2}{x}\,u_0
v_0+(2-3i\delta-\delta^2)\Phi_x^2\right]v_0\Phi^{-3+ i\delta}=0
\end{eqnarray}
\end{subequations}
and  required that their sum vanish. These equations imply that
we must have
\begin{equation}\label{}
u_0v_0=A\frac{\Phi^2}{x}, \qquad \Phi(x,t)=\phi_0(t)\,x^B
\end{equation}
where
\begin{equation}\label{AB}
A=-\frac{3\delta g_1+(\delta^2-2)g_2}{3\delta
q_1+(\delta^2-2)q_2}, \qquad B^2=\frac{q_1A^2+g_1A}{\delta^2-2}.
\end{equation}
If we proceed and consider that coefficients of $\Phi^{-2\mp
i\delta}$ and $\Phi^{-1\mp i\delta}$ vanish, we see that we need
to have   $B=2/3$, $h_1=5/36$, $h_2=0$ and we are able to
determine the form of $u_0$, $v_0$ and $\phi_0(t)$ successfully as follows.
\begin{subequations}\label{}
\begin{eqnarray}
\Phi(x,t)&=&\frac{A^{-1/2}\,x^{2/3}}{(c_1t+c_2)^{2/3}},\\
u_0(x,t)&=&\frac{x^{1/6}}{(c_1t+c_2)^{2/3}}\exp\left[i\frac{c_1 x^2}{4(c_1t+c_2)}\right],\\
v_0(x,t)&=&\frac{x^{1/6}}{(c_1t+c_2)^{2/3}}\exp\left[-i\frac{c_1
x^2}{4(c_1t+c_2)}\right].
\end{eqnarray}
\end{subequations}
We need to re-formulate \eqref{AB} using the condition $B=2/3$. We
skip the details here and sum up the main result.
\begin{prop} The function
\begin{equation}\label{solprop3}
u(x,t)=\sqrt{\frac{A}{x}}\exp\left\{i\left[\frac{c_1x^2}{4(c_1t+c_2)}-\delta\ln\left(\frac{k\,x^{2/3}}{(c_1t+c_2)^{2/3}}\right)\right]\right\}
\end{equation}
solves \eqref{exact} for $h_1=5/36$, $h_2=0$ where
$\delta=-\frac{3A}{4}(g_2+q_2A)$ and $A$ is a root of the
equation
\begin{equation}
32-9A(-4g_1+A(-4q_1+(g_2+q_2A)^2))=0.
\end{equation}
\end{prop}
\textbf{Remark 3.} Proposition 1 also presents a solution to  Eq. \eqref{exact}. \eqref{solprop1i} seems to contain  \eqref{solprop3} when the magnitude is considered. However, Proposition 3 does not have the condition $g_2=q_2=0$, therefore \eqref{solprop3} is valid for a more general choice of the parameters. Furthermore, the solution which is the main theme of \cite{GHO2013}, in which the cubic equation is considered,  coincides with \eqref{solprop3} in special cases.

 We tried the same approach for the canonical equations of $L_2,L_3,L_4$ but this ended up without any success.

\section{Blow-up in $L_p$, $L_\infty$ Spaces  and in Distributional Sense}
The machinery of this section will follow that of \cite{GHO2013}. The main consideration here is going to be the solution \eqref{solprop1i} of Proposition 1 found for the canonical equation  \eqref{eq}. Let us stress what kind of importance  \eqref{eq} has by the following proposition.
\begin{prop}
Any PDE of the form \eqref{canonic} (or, equivalently \eqref{cqsch}) which admits a four-dimensional Lie symmetry algebra containing $\Sl(2,\mathbb{R})$ as a subalgebra can be transformed by point transformations to the canonical form \eqref{eq}. The symmetry algebra of \eqref{eq} is spanned by the vector fields $T,D_1,C_1,W$ noted in \eqref{T}-\eqref{C1}.
\end{prop}
These vector fields produce point transformations of translation in time, scaling, conformal and gauge transformations, respectively. By exponentiating the generators
$T,D_1$ and $C_1$  and combining them we find the $\SL(2,\mathbb{R})$ group action on the solutions to \eqref{eq}.  We state this result in the following Proposition.
\begin{prop}
Given a solution $u_s(x,t)$ of \eqref{eq},
\begin{equation}
 u(x,t)=(a+bt)^{-1/2}\, \mathrm{e}^{{i}\frac{bx^2}{4(a+bt)}}
\,u_s\left(\frac{x}{a+bt},\frac{c+dt}{a+bt}\right)
\end{equation}
is also a solution to \eqref{eq} for $ad-bc=1$.
\end{prop}
Now we take the solution $u_s(x,t)$ to \eqref{eq}  as $\zeta(x)$ defined by \eqref{eq}: $u_s(x,t)=\zeta(x)$ of \eqref{eq}. If we use  Proposition 5, we have that
\begin{equation}\label{uxt}
 u(x,t)=(a+bt)^{-1/2}\, \mathrm{e}^{{i}\frac{bx^2}{4(a+bt)}}
\,\zeta\left(\frac{x}{a+bt}\right)
\end{equation}
also solves \eqref{eq}. Suppose $a>0$, $b<0$ and define $\epsilon=a+bt=b(t-T)$, where $\displaystyle T=-\frac{a}{b}>0$. Let us note that $t\rightarrow T^{-}$ if and only if $\epsilon \rightarrow 0^+$. This enables us to simplify \eqref{uxt} in the form
\begin{equation}\label{ep}
\zeta_\epsilon(x)=\epsilon^{-1/2}\mathrm{e}^{{i}\frac{bx^2}{4\epsilon}}
\,\zeta\left(\frac{x}{\epsilon}\right).
\end{equation}
Let us write down $\zeta_\epsilon(x)$ explicitly:
\begin{equation}\label{zetaepsilon}
\zeta_\epsilon(x)=A\left[\frac{x^\gamma}{k_1 x^{1+\gamma}+k_2\epsilon^{1+\gamma}}\right]^{1/2}\mathrm{e}^{i\left(\frac{bx^2}{4\epsilon}+P_0\right)}.
\end{equation}
Some facts for later convenience: This expression has no discontinuity  for  $x\geq 0$ if $\gamma, k_1,k_2$ and $\epsilon$ are positive. But $x^\gamma$ or $x^{1+\gamma}$ may not be defined for $x<0$. However, this can be prevented by putting the condition that $ \gamma = 2\frac{n_1}{n_2}-1>0$ where $n_1,n_2$ are natural numbers relatively prime. In this case, the solution for $x>0$ and the solution for $x<0$ are the same in modulus but they have a phase difference of $ \frac{\gamma\pi}{2}$. With this condition, $|\zeta_\epsilon|$ becomes an even function. On the other hand, if we do not want to put this condition on $\gamma$, then we may consider the function
\begin{equation}
\tilde{\zeta}_\epsilon(x)=
\begin{cases}
      \zeta_\epsilon(x), & x\geq 0 \\
      0, & x<0. \\
      \end{cases}
\end{equation}
Let us call $\zeta_{\epsilon}(x)$ or $\tilde{\zeta}_\epsilon(x)$  as $u_{\epsilon}(x)$. In case any question arises on whether $u_\epsilon$ is indeed a solution in regard to differentiability issues at $x=0$: By a direct calculation we see that $u_{\epsilon}(x)$  is twice differentiable with respect  to $x$  at $x=0$ for $\epsilon>0$ if $\gamma>4$ (the value of the second derivative being zero). Therefore, $u_\epsilon(x)$ is a classical solution to \eqref{eq} on $-\infty < x < \infty$, $t<T$ for $\gamma>4$.  Observe that
\begin{equation}
\int_{-\infty}^\infty |u_\epsilon(x)|dx=\sigma \int_{0}^\infty |u_\epsilon(x)|dx
\end{equation}
where $\sigma=2$ if $u_\epsilon=\zeta_\epsilon$ and  $\sigma=1$ if  $u_\epsilon=\tilde{\zeta}_\epsilon$.

\subsection{Blow-up in $L_p(\mathbb{R})$}
\begin{thm}
For any $T>0$ there exists a solution $u(x,t)$ to Eq. \eqref{eq} such that
\begin{equation}\label{}
\lim_{t\rightarrow T^{-}}\|u(x,t)\|_p=\infty, \qquad p>2
\end{equation}
where $\displaystyle\|u(x,t)\|_p=\Big(\int_{-\infty}^{\infty}|u|^pdx\Big)^{1/p}$.
\end{thm}
\textbf{Proof.} Let $T>0$ be given. We will take $u(x,t)$ as $u_\epsilon(x)$ we have defined. We are free to choose numbers $a,b$ such that $a>0$, $b<0$ and $\displaystyle T=-\frac{a}{b}$. We can say that
\begin{equation}\label{}
\lim_{t\rightarrow T^{-}}\|u(x,t)\|_p=\lim_{\epsilon \rightarrow 0^+}\|u_\epsilon(x)\|_p.
\end{equation}
Let us calculate $L_p(\mathbb{R})$ norm of $u_\epsilon(x)$.
\begin{equation}\label{}
||u_\epsilon(x)||_p^p=\sigma|A|^p\int_{0}^\infty\Big[\frac{x^\gamma}{k_1 x^{1+\gamma}+k_2\epsilon^{1+\gamma}}\Big]^{p/2}dx.
\end{equation}
If we substitute $x=\epsilon y$, we get
\begin{align}\label{}
||u_\epsilon(x)||_p^p&=\frac{\sigma|A|^p}{\epsilon^{\frac{p-2}{2}}}\int_{0}^\infty\Big[\frac{y^\gamma}{k_1y^{1+\gamma}+k_2}\Big]^{p/2}dy\nonumber\\
                     &=\frac{\sigma|A|^p}{\epsilon^{\frac{p-2}{2}}}\left\{\int_{0}^1\Big[\frac{y^\gamma}{k_1 y^{1+\gamma}+k_2}\Big]^{p/2}dy+ \int_{1}^\infty\Big[\frac{y^\gamma}{k_1 y^{1+\gamma}+k_2}\Big]^{p/2}dy\right\}.
\end{align}
Clearly, the first integral is finite due to the continuity of the integrand and the second integral is finite iff $p>2$. Therefore for $p>2$ we get
\begin{equation}\label{}
\lim_{\epsilon\rightarrow 0^+}||u_\epsilon(x)||_p=\frac{\sigma^{1/p}|A|}{\epsilon^{\frac{p-2}{2p}}}\left(\int_{0}^\infty\Big[\frac{y^\gamma}{k_1y^{1+\gamma}+k_2}\Big]^{p/2}dy\right)^{1/p}=\infty.\\
\end{equation}
This completes the proof of Theorem 1.
\subsection{Blow-up in $L_\infty(\mathbb{R})$}
The following theorem concerns  $u_\epsilon(x)$ and shows that it blows up in $L_\infty(\mathbb{R})$ norm.
\begin{thm}
Given any $T>0$, there exists a solution to \eqref{eq} such that
\begin{equation}\label{}
\lim_{t\rightarrow T^{-}}\|u(x,t)\|_\infty=\infty.
\end{equation}
Here $\|u(x,t)\|_\infty = \mathrm{ess}\sup_{x\in [0,\infty)}|u(x,t)|$, $t<T$.
\end{thm}
\textbf{Proof.} We again concentrate on the solution  $u_\epsilon(x)$, for which we similarly write
\begin{equation}\label{}
\lim_{t\rightarrow T^{-}}\|u(x,t)\|_\infty=\lim_{\epsilon \rightarrow 0^+}\|u_\epsilon(x)\|_\infty.
\end{equation}
By the definition of  $u_\epsilon(x)$ we can restrict our attention to the interval $[0,\infty)$. Since $|u_\epsilon(x)|$ is continuous on $[0,\infty)$,
$|u_\epsilon(0)|=0$ and $\lim_{x\rightarrow \infty}|u_\epsilon(x)|=0$, there exists $x_0\in (0,\infty)$ such that
\begin{equation}\label{}
||u_\epsilon(x)||_\infty=\mathrm{ess}\sup_{x\in [0,\infty)}|u_\epsilon(x)|=\max_{x\in [0,\infty)} |u_\epsilon(x)|=|A|\left|\frac{x_0^\gamma}{k_1 x_0^{1+\gamma}+k_2\epsilon^{1+\gamma}}\right|^{1/2}.
\end{equation}
We find that this maximum occurs at
\begin{equation}\label{x0}
\displaystyle x_0=\epsilon\Big(\frac{k_2\gamma}{k_1}\Big)^{\frac{1}{1+\gamma}}
\end{equation} and
\begin{equation}\label{}
||u_\epsilon(x)||_\infty=\frac{K_0}{\sqrt{\epsilon}}, \qquad K_0^2=\frac{\gamma^\frac{\gamma}{1+\gamma}}{k_1^{\frac{\gamma}{1+\gamma}}k_2^{\frac{1}{1+\gamma}}(1+\gamma)}. \end{equation}
Therefore we arrive at
\begin{equation}\label{}
\lim_{\epsilon \rightarrow 0^+}\|u_\epsilon(x)\|_\infty=\infty.
\end{equation}
Before the conformal transformation, the solution \eqref{solprop1i} is a stationary solution. The conformal transformation gives the time-dependent solution \eqref{zetaepsilon}. The stationary solution is obtained by putting $\epsilon=1$. By the action of the conformal transformation the maximum of the stationary solution at the point $x=\Big(\frac{k_2\gamma}{k_1}\Big)^{\frac{1}{1+\gamma}}$ becomes a moving one as it is in \eqref{x0} and the maximum point approaches the origin as $\epsilon\rightarrow 0^+ \iff t\rightarrow T^-$, whilst the maximum value inreases unuboundedly. This is the case if $u_\epsilon=\tilde\zeta_\epsilon$. If $u_\epsilon=\zeta_\epsilon$ there are two peaks at $\mp x_0$ approaching origin and blowing up.
\subsection{Blow-up in the space of generalized functions}
We denote the space of infinitely differentiable functions with compact support by $D$ and $D'$ is the dual of $D$, the space of generalized functions.

\begin{thm}
When $p>2$,
\begin{equation}
\epsilon^{\frac{p-2}{2}}|u_\epsilon (x)|^p\longrightarrow K \delta(x)  \quad \text{as}  \quad \epsilon \rightarrow 0^+
\end{equation}
in $D'$ where $\displaystyle K=\int_{0}^\infty \sigma |A|^p\Big|\frac{y^\gamma}{k_1y^{1+\gamma}+k_2}\Big|^{p/2}dy$ and $\delta(x)$ denotes the Dirac distribution at the origin.
\end{thm}
\textbf{Proof.} We need to show that
\begin{equation}
\lim_{\epsilon \rightarrow 0^+}\int_{-\infty}^\infty \epsilon^{\frac{p-2}{2}}|u_\epsilon (x)|^p \varphi(x)dx=K\varphi(0), \quad \forall \varphi \in D.
\end{equation}
Let us denote the integral on the left by $I(\epsilon)$.
\begin{equation}
I(\epsilon)=\int_{-\infty}^\infty \epsilon^{\frac{p-2}{2}}|u_\epsilon (x)|^p \varphi(x)dx=\epsilon^{\frac{p-2}{2}}\sigma|A|^p\int_{0}^\infty\left|\frac{x^\gamma}{k_1 x^{1+\gamma}+k_2\epsilon^{1+\gamma}}\right|^{p/2}\varphi(x)dx.
\end{equation}
Plugging in $x=\epsilon y$ gives
\begin{equation}
I(\epsilon)=\int_{0}^\infty \sigma |A|^p\Big|\frac{y^\gamma}{k_1y^{1+\gamma}+k_2}\Big|^{p/2}\varphi(\epsilon y)dy.
\end{equation}
Let $\displaystyle \sigma |A|^p\Big|\frac{y^\gamma}{k_1y^{1+\gamma}+k_2}\Big|^{p/2}\varphi(\epsilon y)=f_\epsilon(y)$. Since $\varphi$ is of compact support,  we have
\begin{align}
& |f_\epsilon(y)|\leq C_\epsilon\sigma |A|^p\Big|\frac{y^\gamma}{k_1y^{1+\gamma}+k_2}\Big|^{p/2}   \in L_1([0,\infty)) \quad   \text{if} \quad  p>2, \\
& \lim_{\epsilon \rightarrow 0^+} f_\epsilon(y)=\sigma|A|^p\Big|\frac{y^\gamma}{k_1y^{1+\gamma}+k_2}\Big|^{p/2}\varphi(0).
\end{align}
By Lebesgue's dominated convergence theorem we find that
\begin{equation}
\lim_{\epsilon \rightarrow 0^+} \int_{0}^\infty f_\epsilon(y)dy=\varphi(0)\int_{0}^\infty \sigma |A|^p\Big|\frac{y^\gamma}{k_1y^{1+\gamma}+k_2}\Big|^{p/2}dy=K\varphi(0),
\end{equation}
which implies
\begin{equation}
\epsilon^{\frac{p-2}{2}}|u_\epsilon (x)|^p\longrightarrow K \delta(x)
\end{equation}
as $\epsilon \rightarrow 0^+$.

Let us finally stress that a similar machinery for blow-up in distributional sense is carried out in \cite{Ozawa1992,EEM2006} for Davey-Stewartson type equations.

\section{Conclusion}

In this paper we analyzed a class of variable coefficient cubic-quintic nonlinear Schr\"{o}dinger equations. We concentrated on representative equations of families with
Lie symmetry algebras of dimension four. What is special about these equations is that they cannot be converted to standard constant-coefficient cubic-quintic NLS equation. They do have variable coefficients. In a methodological way, we obtained several different reductions of these PDEs to ODEs. Among them for the ones  being Painlevé-integrable, we referred to results existing in literature.  We have seen that many of the reduced ODEs inherited the non-integrable character of the relevant PDE, leaving us with less chance for presenting exact analytical solutions. However, we still have been successful for doing so, using truncated expansions of Painlevé series.

The truncated expansions were tried in obtaining exact solutions both for  each of the four canonical PDEs themselves and for their seven second order reductions. The attack was fruitful for two of the second-order reductions by producing three different solutions, and for one of the PDEs itself by producing one exact solution.

That being not the end of the story, we used one of these exact solutions to expose the blow-up character of solutions of one of the canonical PDEs, which includes inhomogeneous nonlinearities and an inhomogeneous potential. The blow-up character is due to conformal invariance of the equation, and occurs in $L_p$ norm of the solution with $p>2$, in $L_\infty$ norm and in distributional sense. All solutions we have considered are to be taken in the classical sense.

We would like to finalize by noting how the results of this paper may be of (further) use. We have presented some exact solutions for variable-coefficient PDEs. They will be useful when the PDE is to model a real phenomenon. If not at all, these solutions can be used for the convergence test of numerical algorithms of PDEs. We have obtained many reduced ODEs of second and third order which did not have the Painlevé property therefore not giving us the chance to integrate them by using the known results in literature for P-integrable equations. Being born on earth from a well-known equation of mathematical physics, these equations may be analyzed for possible classical and non-classical symmetries (e.g. $\lambda$-symmetries). What is more; these reduced ODEs may also be useful in cases where the numerical schemes for the PDE they were obtained require high computational cost or produce divergences, as there are many sophisticated  numerical algorithms  for the treatment of an ODE \cite{Muzo}. Finally,  to our best knowledge, the search for formation of blow-up in suitable Sobolev spaces remains as an open problem.

\section*{Acknowledgements}
I would like to thank Prof. Faruk G\"{u}ng\"{o}r for his valuable comments on the progress of this paper and on reading the manuscript. The  results of Painlevé tests mentioned above are owed to a  package developed by D. Baldwin and W. Hereman \cite{Hereman}.

\end{document}